\documentclass[a4paper,11pt]{article}
\usepackage{pos}
\begin{document}
\title{From tensors to qubits}
\ShortTitle{From tensors to qubits}

\author*[a]{Yannick Meurice}\author[b,c]{Alexei Bazavov}
\author[d]{Patrick Dreher}
\author[a,e]{Erik Gustafson}
\author[c,b]{Leon Hostetler}
\author[a]{Ryo Sakai}
\author[f]{Shan-Wen Tsai}
\author[e]{Judah Unmuth-Yockey}
\author[a]{Jin Zhang}

\affiliation[a]{Department of Physics and Astronomy, The University of Iowa\\
Iowa City, IA 52242, USA}

\affiliation[b]{Department of Computational Mathematics, Science and Engineering, Michigan State University, \\East Lansing, Michigan 48824, USA}

\affiliation[c]{Department of Physics and Astronomy, Michigan State University, East Lansing, Michigan 48824, USA}

\affiliation[d]{Electrical and Computer Engineering, North Carolina State University,\\
  Raleigh, NC  27695, USA}

\affiliation[e]{Fermilab, Batavia, IL 60510, USA}

\affiliation[f]{Department of Physics and Astronomy, UC Riverside, Riverside, CA 92521, USA}


\emailAdd{yannick-meurice@uiowa.edu}



\abstract{We discuss recent progress in Tensor Lattice Field Theory and economical, symmetry preserving, truncations suitable for quantum computations/simulations. We focus on spin and gauge models with continuous Abelian symmetries such as the Abelian Higgs model and emphasize noise-robust implementations of Gauss’s law. We discuss recent progress concerning the comparison between field digitizations and character expansions, symmetry breaking in tensor language, wave-packet preparation and possible new implementations of Abelian models using Rydberg atoms.}

\FullConference{%
 The 38th International Symposium on Lattice Field Theory, LATTICE2021
  26th-30th July, 2021
  Zoom/Gather@Massachusetts Institute of Technology
}


\maketitle

\def\bpl{\beta_{pl.}}
\def\bl{\beta_{l.}}

\def\zq{${\mathbb Z}_q\ $}

\def\te{${ \mathbb T}_E\ $}
\def\tm{${ \mathbb T}_M\ $}

\def\vone{{\mathbb V}_1 }
\def\vtwo{{\mathbb V}_2 }

\def\mc{\mathcal{C}}
\def\mcm{\mathcal{C}^{-1}}
\def\np{n_{+1}}
\def\nm{n_{-1}}
\def\uo{\frac{U}{2}}
\def\no{n_0}
\def\do{\Delta _0}
\def\vo{V_0}
\def\vop{\frac{V_0}{64}}
\graphicspath{{./figures/}}
\def\beq{\begin{equation*}}
\def\enq{\end{equation*}}
\def\bra#1{\mathinner{\langle{#1}|}}
\def\ket#1{\mathinner{|{#1}\rangle}}
\def\braket#1{\mathinner{\langle{#1}\rangle}}
\def\Bra#1{\left\langle#1\right|}
\def\Ket#1{\left|#1\right\rangle}
\def\lt{\lambda ^t}
\def\note{note}
\def\beq{\begin{equation}}
\def\enq{\end{equation}}
\def\hata{\hat{\alpha}}                                                                    
\def\hatx{\hat{x}}                                                                    
\def\hatt{\hat{t}}                                                                    
\def\ms{\text{-}s}                                                                    
\def\mt{\text{-}t}                                                                    
\def\Tr{{\rm Tr}\,}

\def\EE{entanglement entropy }
\def\TE{thermal entropy }

\titlepage


\section{Introduction}

Tensor network methods are playing an increasingly important role in lattice gauge theory. 
The annual lattice conferences certainly helped fostering contacts between the communities involved 
in the early developments \cite{mcbpos2013,ympos2013,kvapos2014,yspos2014,mcbpos2014,juypos2014,ympos2014}. 
The number of contributions has grown steadily with the years. In 2021, the keyword ``tensor network" appears 24 times in the book of abstracts and the keyword ``tensor renormalization" 12 times. 

The tensor network approaches can be divided roughly into two categories. The first one relies on Hamiltonian formulations and often makes use of powerful computational tools developed in the condensed matter community, such as the Matrix Product States (MPS) in 1+1 dimensions and higher-dimensional extensions such as 
Projected Entangled Pair States (PEPS). For recent general reviews see \cite{hv,ran}; for reviews focused on lattice gauge theory see \cite{mcbpos2018,Banuls:2019bmf}; for recent use of PEPS in gauge theories see \cite{emonts}. For Hamiltonian formulations of non-Abelian gauge theories at this conference, see \cite{davoudi}. For Hamiltonian studies of the Abelian Higgs and $O(2)$ models at this conference see \cite{jz,swt}.  

The other category of approaches is based on the conventional Lagrangian lattice formulation and relies on various types of dualities and character expansions. It is sometimes called Tensor Lattice Field Theory (TLFT). For recent reviews that provide a roadmap resembling the one followed by standard review articles \cite{kogut79,kogut83} on lattice gauge theory (the ``Kogut sequence") see \cite{rmp,book}. When the field variables are  continuous and compact, character expansions provide discrete indices that fit naturally the needs of quantum computing. This is the main theme of these proceedings. 

TLFT provides reformulations that are exactly equivalent to the original lattice models. 
One of the original motivations for this approach was the possibility of doing {\it local} coarse-graining \cite{exact,exact2,rmp}. 
This involves approximations (``truncations") that can be checked in regimes 
where the usual Monte Carlo sampling in configuration space is possible. 
This is often called the Tensor Renormalization Group (TRG).
For examples in this conference see for instance \cite{dk,akipos2021,blochpos2021,rs,leonpro}. Another possibility is to sample the tensor configurations appearing in the reformulations as in the worm algorithm \cite{worm,wolff,worm2,ympos2014,worm3,worm4}. These approaches are completely Lagrangian based. However, it is possible to connect with the Hamiltonian approach by using the transfer matrix and taking the time-continuum limit \cite{rmp,book,hv,worm3}. 

Basic aspects of TLFT are reviewed in Sec. \ref{sec:tlgt}. Because of the equivalence with the standard formulation, symmetries of the action lead to known identities or theorems which need to 
be recast in the new formulation. In particular, local gauge invariance is manifest at every step of the reformulation. In addition, truncations preserve symmetries \cite{ymex,ymdis}. These questions and economical implementation of Gauss's laws are briefly discussed in Sec. \ref{sec:trunc}. The simple case of the Abelian Higgs model in 1+1 dimension and a possible implementation with Rydberg-dressed atoms in optical lattice \cite{tahm,ploop} are reviewed in Sec. \ref{sec:optical}. This inspired proposals to implement the same model with configurable arrays of Rydberg atoms \cite{51atom,keesling} which were presented at the conference. This is discussed in Sec. \ref{sec:rydberg}. More details can be found in a recent publication \cite{ymryd}. 
Finally, related results presented at this conference \cite{pdr,leonpro,swt,jz} are briefly summarized in Sec. \ref{sec:related}.

\section{Tensor Lattice Field Theory  for Quantum Computing}
\label{sec:tlgt}
Quantum computing for quantum field theory requires a complete discretization. 
Discretization of space is well understood from the success of lattice QCD.
Discretization of field integration for continuous field variables can be achieved with various methods. In most lattice simulations, the variables of integration are compact and tensor formulations use character expansions (such as Fourier series) to re-express the partition function as discrete sums of contracted tensors.
The ``hard" integrals are done exactly and field integrations provide Kronecker deltas that encode the symmetries.
For continuous field variables, the sums are infinite, but truncations to finite sums do not break symmetries \cite{ymex,ymdis}. Our general expectation is  that the correct large distance physics can be obtained with ``a few qubits per unit cells". Comparison with other 
methods based on field discretization or quantum links are highly desirable \cite{Klco:2018zqz,Lamm:2019bik,Brower:2020huh,worm4}. 

As a simple illustration of the discrete reformulation, we consider the lattice non-linear $O(2)$ model. 
Its partition function reads:
\beq
Z_{O(2)}=\prod_x\int_{-\pi}^{\pi}\frac{d\varphi _x}{2\pi}e^{\beta \sum\limits_{x,\mu} \cos(\varphi_{x+\hat{\mu}}-\varphi_x)}.\enq
Using the character expansion, which in this case amounts to a Fourier series,  we expand the Boltzmann weights as 
\beq
   e^{\beta \cos(\varphi_{x+\hat{\mu}} - \varphi_{x})} = \sum_{n_{x,\mu}=-\infty}^{\infty} 
   e^{i n_{x,\mu} (\varphi_{x+\hat{\mu}}-\varphi_x)} I_{n_{x,\mu}}(\beta),
\enq
and obtain
\beq
Z_{O(2)}={\rm Tr} \prod_x T^{(x)}_{n_{x-\hat{1},1}, n_{x,1},\dots,n_{x,D}}.
\enq
with a {\it local} tensor $T^{(x)}$ with $2D$ indices. The explicit form is
\beq
  \label{eq:tensor}
  T^{(x)}_{n_{x-\hat{1},1}, n_{x,1},\dots,n_{x-\hat{D},D},n_{x,D}}=
  \sqrt{I_{n_{x-\hat{1},1}} I_{n_{x,1}},\dots, I_{n_{x-\hat{D},D}} I_{n_{x,D}}  }\times\delta_{n_{x, \text{out}},n_{x, \text{in}}},
\enq
Schematically, we replace integrations over circles located at the sites by discrete sums over indices located on links: 
\beq
\textcolor{blue}{\prod_x\int_{-\pi}^{\pi}d\varphi _x} \Longrightarrow  \textcolor{red}{\sum_{\{ n \} }}
\enq
This can be ilustrated graphically for $D=2$ as:  
\begin{figure}[h]
\centering
\includegraphics[width=7cm,angle=0]{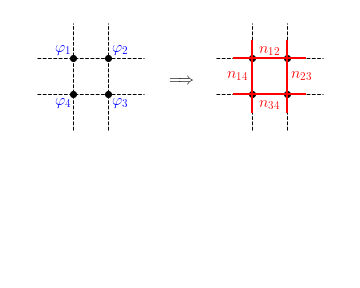}
\vskip-3cm
\end{figure}

\section{Truncations, symmetries and Gauss's law}
\label{sec:trunc}

Tensorial truncations are compatible with the general identities derived from global and local symmetries \cite{ymex,ymdis}. 
Symmetries are encoded in tensor selection rules and {\it not} in the numerical values of the tensor elements. 
Universal properties of these models can be reproduced with highly simplified formulations desirable for implementations with quantum computers or for quantum simulations experiments ("few qubits per unit cells").
In tensorial reformulations Noether theorem can be stated as: for each symmetry there is a corresponding tensor redundancy (and we can "fix" the corresponding integration variable). This applies not only to continuous but also discrete symmetries. Note however, that truncations may affect the type of phase transitions \cite{jin,jz,swt}.

Noise robust enforcement of Gauss's law is easy for bosonic matter.  For the Abelian Higgs model, the gauge quantum numbers determine the matter quantum numbers completely. This is a discrete version of $\partial _j E^j=\rho$. In the pure gauge case, one can associate new quantum numbers with the space-space plaquettes \cite{ymdis} in a discrete version of 
$E^k=\partial _j C^{jk}$. See also \cite{Bender:2020ztu,Haase:2020kaj,Stryker:2020sap,juydual} for related discussions.

\section{Optical lattice implementation of the Abelian Higgs model}
\label{sec:optical}
One of the simplest non-trivial gauge theories is the Abelian Higgs model. See \cite{tahm} for the TLFT formulation. 
After decoupling the Higgs model and taking the time-continuum limit in 1+1 dimensions, we obtain an Hamiltonian that can be implemented on an optical lattice \cite{ploop} as illustrated in the figure below for a spin-2 approximation.  
\beq
H=\textcolor{red}{\frac{U}{2}\sum_i \left(L^z_{(i)}\right)^2}+ \textcolor{blue}{\frac{Y}{2}\sum_i (L^z_{(i)}-L^z_{(i+1)})^2}-\textcolor{green}{
X\sum_{i}
 U^x_{(i)}}.\enq
\begin{figure}[h]
 \centering
 \vskip-5cm
\includegraphics[width=12cm,angle=0]{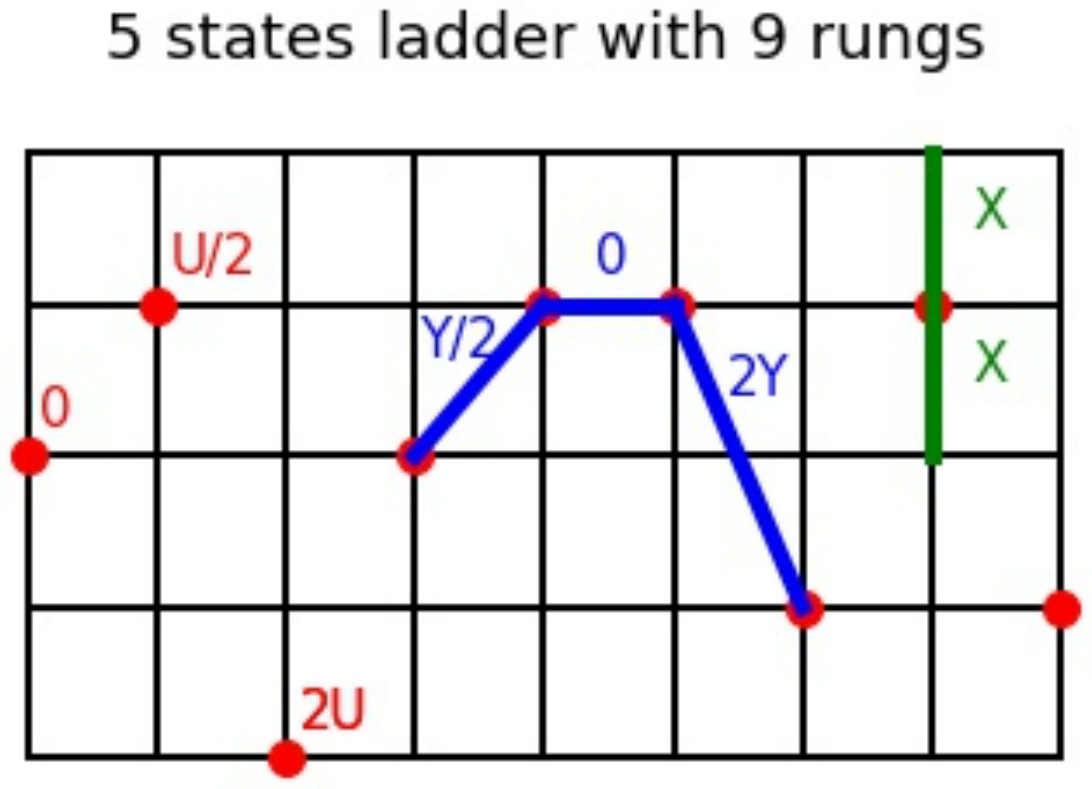}
\vskip-5cm
\end{figure}
\vskip-0.1cm
\noindent
The color coded  explanations \cite{book} for an optical lattice ladder with one atom per rung are: tunneling along the vertical direction ($\bar{L}^z=\pm2, \ \pm1,\ 0$, \textcolor{green}{green}), no tunneling in the the horizontal direction but short range attractive (Rydberg-dressed) interactions (\textcolor{blue}{blue}). A parabolic potential is applied in the spin (vertical) direction (\textcolor{red}{red}). See \cite{tahm,ploop} for details.

\section{Rydberg atom simulators}
\label{sec:rydberg}
Arrays of $^{87}Rb$ atoms separated by adjustable  distances, homogeneously coupled to the excited Rydberg state $\ket{r}$ with detuning $\Delta$  \cite{51atom} offer new possibilities for quantum simulations \cite{celi,surace,qedryd}. Each atom has 2 states:  $\ket{g}$, the ground state, and $\ket{r}$ the Rydberg state. We use 
$n$ to denote the occupation of $\ket{r}$;  in the ``qubit" picture: $\ket{g}\rightarrow \ket{0}$ and $\ket{r}\rightarrow \ket{1}$.
The Hamiltonian reads
\beq
\label{eq:genryd}
H = \frac{\Omega}{2}\sum_i(\ket{g_i}\bra{r_i} + \ket{r_i}\bra{g_i})-\Delta\sum_i  n_i +\sum_{i<j}\Omega R_b^6/R_{ij}^6n_in_j,
\enq
with $i$ and $j$ run over atoms in configurable arrays.

We will make use of the Rydberg blockade which means that $\ket{rr}$ is suppressed if two atoms are too close.
Programming means assembling atom arrays with tweezers. This setup was used to identify a rich phase diagram in the (distance between atoms, $\Delta/\Omega$) plane \cite{keesling}.

We now turn to proposals for spin-2 and spin-1 truncations presented at the conference and discussed in more details in \cite{ymryd}. 
A simple adaptation of the spin-2 optical lattice construction of Sec. \ref{sec:optical} is illustrated below. As discussed below, the $X$ term can be induced by the $\Omega$ interactions. 
\begin{figure}[h]
\centering
\includegraphics[width=9cm]{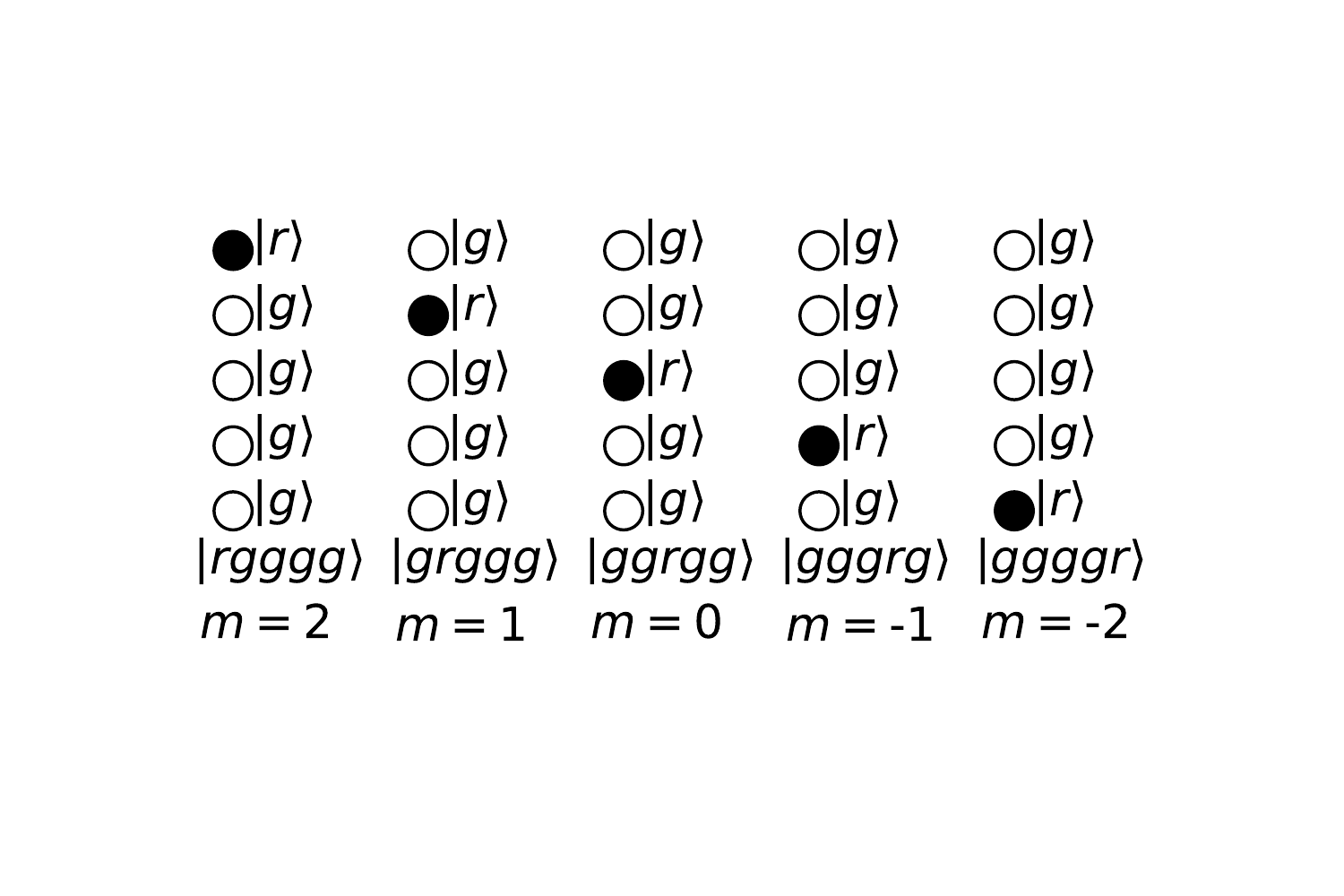}\vskip2pt
\end{figure}
\vskip-1cm

In view of the currently available technology, we suggested to start with a  simpler construction for spin-1: 
\begin{figure}[h]
\centering
\includegraphics[width=8cm]{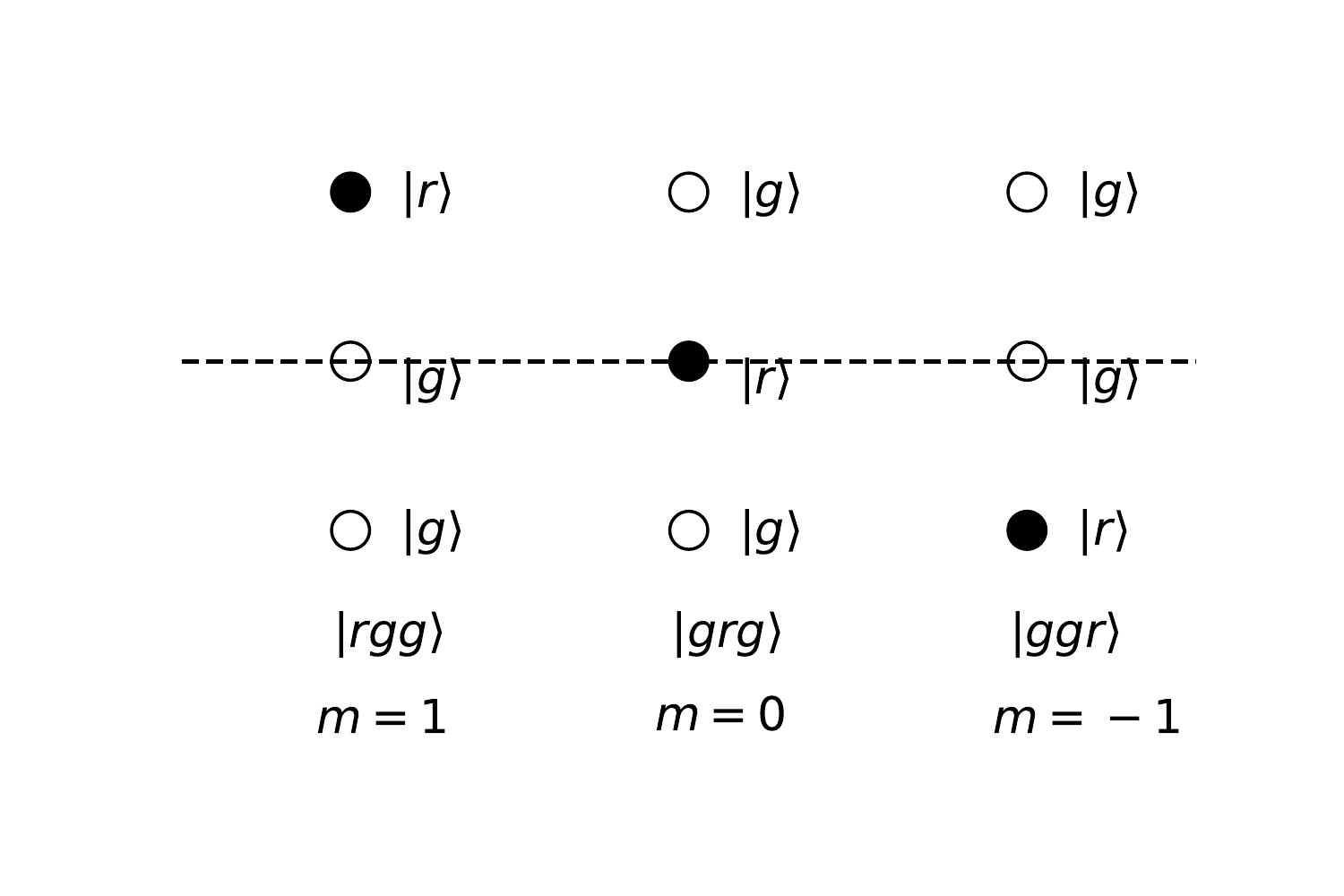}\vskip2pt
\end{figure}
\vskip0.1cm
\noindent
Each set of three atoms can be seen as a qutrit (for motivations see \cite{Gustafson:2021qbt,Ciavarella:2021nmj}). 
The $X$-term (tunneling in the optical lattice formulation) is induced by 
transitions to heavy states at second order in perturbation theory \cite{ymryd}. 

An even more economical version with only two atoms per links is given below:
\begin{figure}[h]
\centering
 \includegraphics[width=9cm,angle=0]{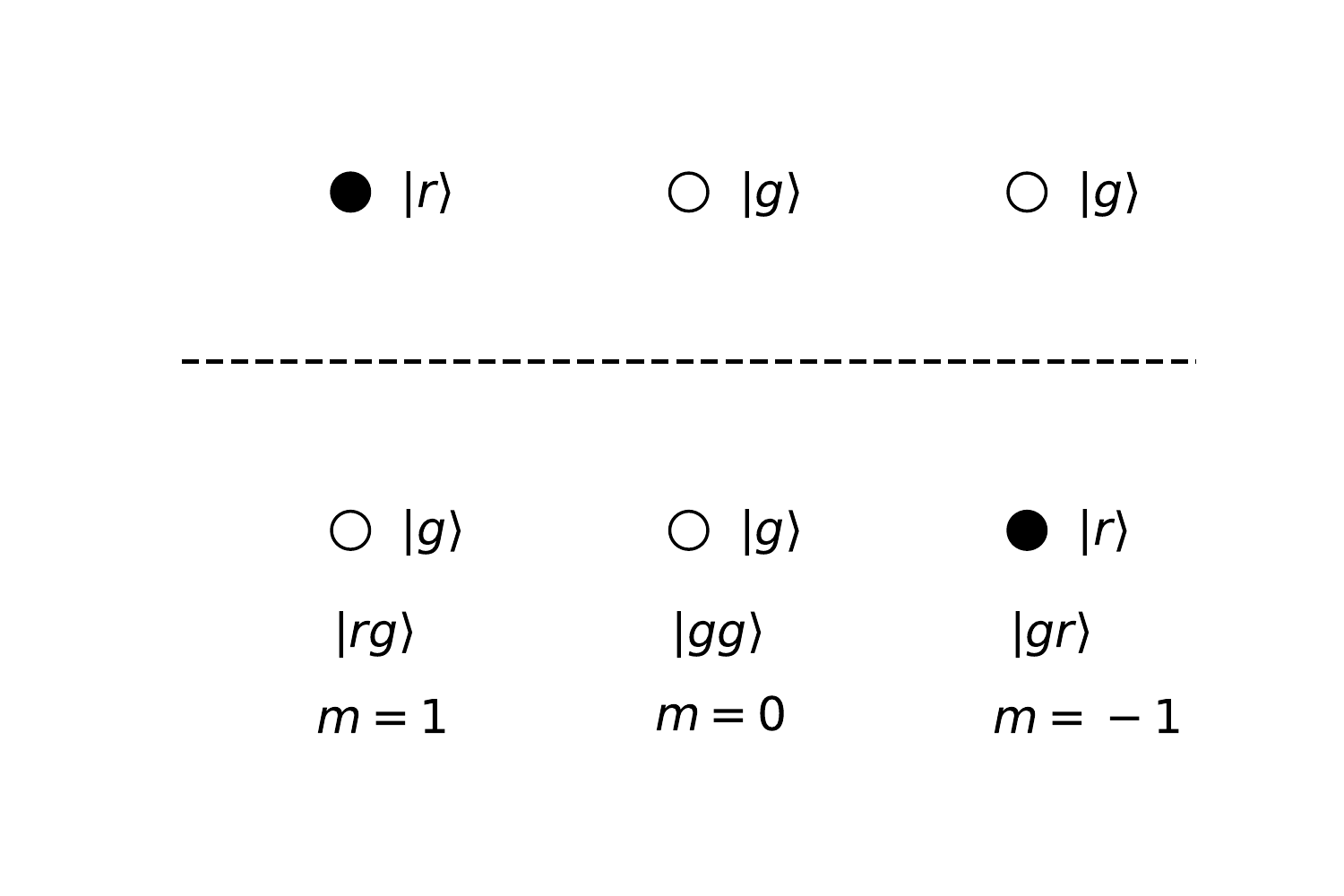}  
\end{figure}

For the one spin system we can compare the simulator  with the target model for spin-1:  
\beq
H=\frac{U}{2}(L^z)^2-XU^x.
\enq
$H$ is invariant under charge conjugation: we use $\mc$ eigenstates
\beq
\ket{\pm}\equiv \frac{1}{\sqrt{2}}(\ket{1}\pm\ket{-1}).
\enq
\beq\mc \ket{\pm}=\pm\ket{\pm}.\enq
They are also eigenstates of $(L^z)^2$ with eigenvalue 1, and 
$\mc \ket{0}=\ket{0}$.
There is only one $\mc$-odd state which is $\ket{-}$.
\beq
U^x\ket{-}=0; \ 
H\ket{-}=\frac{U}{2}\ket{-}.
\enq

The two atom setup discussed above provides a very simple implementation of the one-spin system. Its Hamiltonian reads
\begin{align}
\label{eq:2rah}
H^{2R} = &-\Delta(\np+\nm)+\vo\np\nm \nonumber \\&+\frac{\Omega}{2}\sum_{\pm 1}(\ket{g_{\pm 1}}\bra{r_{\pm 1}} + \ket{r_{\pm 1}}\bra{g_{\pm 1}})\nonumber.
\end{align}
The correspondence between the states of the simulator and the target is given below:
\begin{center}
\begin{tabular}{ |c|c|c|c| } 
\hline
state&ket&target &energy ($\Omega=0$)\\
 \hline
 $\begin{matrix}
\bullet\\
\circ
\end{matrix}$&$\ket{rg}$& $\ket{1}$ & $-\Delta$ \\ 
\hline
 $\begin{matrix}
 \circ\\
\circ
\end{matrix}$&$\ket{gg}$& $\ket{0}$ & 0 \\ 
\hline
 $\begin{matrix}
\circ\\
\bullet
\end{matrix}$&$\ket{gr}$& $\ket{-1}$ & $-\Delta$ \\ 
\hline
 $\begin{matrix}
\bullet\\
\bullet
\end{matrix}$&$\ket{rr}$& - & $-2\Delta+\vo$ \\ 
 \hline
\end{tabular}
\end{center}

\begin{figure}[h]
 \centering
  \includegraphics[width=6.5cm]{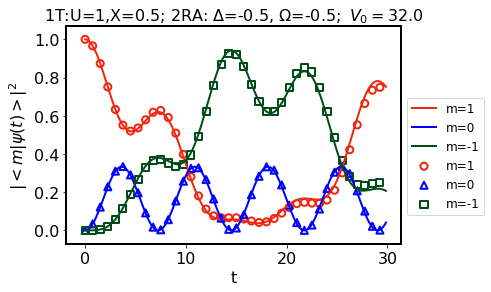}  
  \caption{\label{fig:ex1}$|\bra{m}U(t)\ket{m=1}|^2$, one site with exact Hamiltonian $U=1,\ X=0.5$ (solid lines) and Rydberg Hamiltonian with $\Omega=-0.5$, $\Delta=-0.5$  and $V_0=64|\Omega |=32$ (empty symbols) as in \cite{ymryd}. }
\end{figure}

The matching between the target and the two-atom simulator
is simply $\Delta=-\frac{U}{2}$ and $\Omega=-X$. 
Except for possible transitions to $\ket{rr}$, the correspondence is exact and the linear formula applies for arbitrary values of $X$. A comparison between the target and the simulator is given in Fig. \ref{fig:ex1}.  More details for setups with 3 (the other spin-1 proposal), and 4 or 6  atoms with spin-spin interactions can be found in \cite{ymryd}. 

\section{Related results}
\label{sec:related}
In this section, we give a brief summary of related results presented at this conference. 
\subsection{Phase shifts from real time evolution}
The quantum Ising model provides a nice connection between the TLFT formulation of the Ising model and quantum computing. 
In \cite{pdr}, methods to construct wavepackets and measure phase shifts using IBMQ and trapped ions facilities \cite{phase} were presented.
A renormalized reflection probability that can be measured in the early stages of a collision led to estimates of a time delay related to the phase shift 
by the Wigner formula: 
\beq\Delta t^\star=\delta'(k)/(\partial E/\partial k) . \enq

\subsection{Critical behavior of Spin truncations  in the $O(2)$ limit}
In \cite{jz,swt}, the effect of truncations on the critical behavior were discussed for the $O(2)$ model  and the Abelian Higgs model. 
For instance \cite{jin}, in the case of the charge representation of the $O(2)$ model
\beq
\hat{H}_{charge} = \frac{Y}{2} \sum_{l=1}^{L+1} (\hat{S}_{l}^z)^2 - \frac{X}{2} \sum_{i = 1}^{L} (\hat{U}_{l}^+ \hat{U}_{l+1}^- + \hat{U}_{l}^- \hat{U}_{l+1}^+),\enq
energy gaps  $\Delta E_{V=\infty}$ for spin truncations $S = 1, 2, 3, 4$ were considered. 
For $S \ge 2$ a behavior of the type $A \exp(-b/\sqrt{Y-Y_c})$ as in regular Kosterlitz Thouless (KT) was found to be accurate. However for 
$S = 1$, the parametrization $A \sqrt{Y-Y_c} \exp\left[-b/(Y-Y_c)\right]$  which corresponds to a $SU(2)$ symmetry on the KT separatrix was found to be more accurate.

\subsection{Interpolation among $Z_q$ clock models}
In \cite{leon,leonpro} a $O(2)$ model with symmetry  breaking : 
\beq \Delta S=\gamma\sum_x\cos(q\varphi_x), \enq was considered. In the limit of large $\gamma$, the angles $\varphi=\frac{2\pi k}{q}\ k=0,1, ..,\lfloor q \rfloor$ are strongly favored. The study of the symmetry breaking in tensor language is quite interesting and give accurate results at small $\beta$. For 
integer $q$, we have a  $Z_q$ symmetry but for 
non-integer $q$ there is only a $Z_2$ symmetry.
The schematic phase diagram from \cite{leon} is shown below. The finite $\gamma$ region remains to be explored and could be related to the  Rydberg arrays phase diagram in \cite{keesling}. 
\begin{figure}[h]
\centering
 \includegraphics[width=7cm]{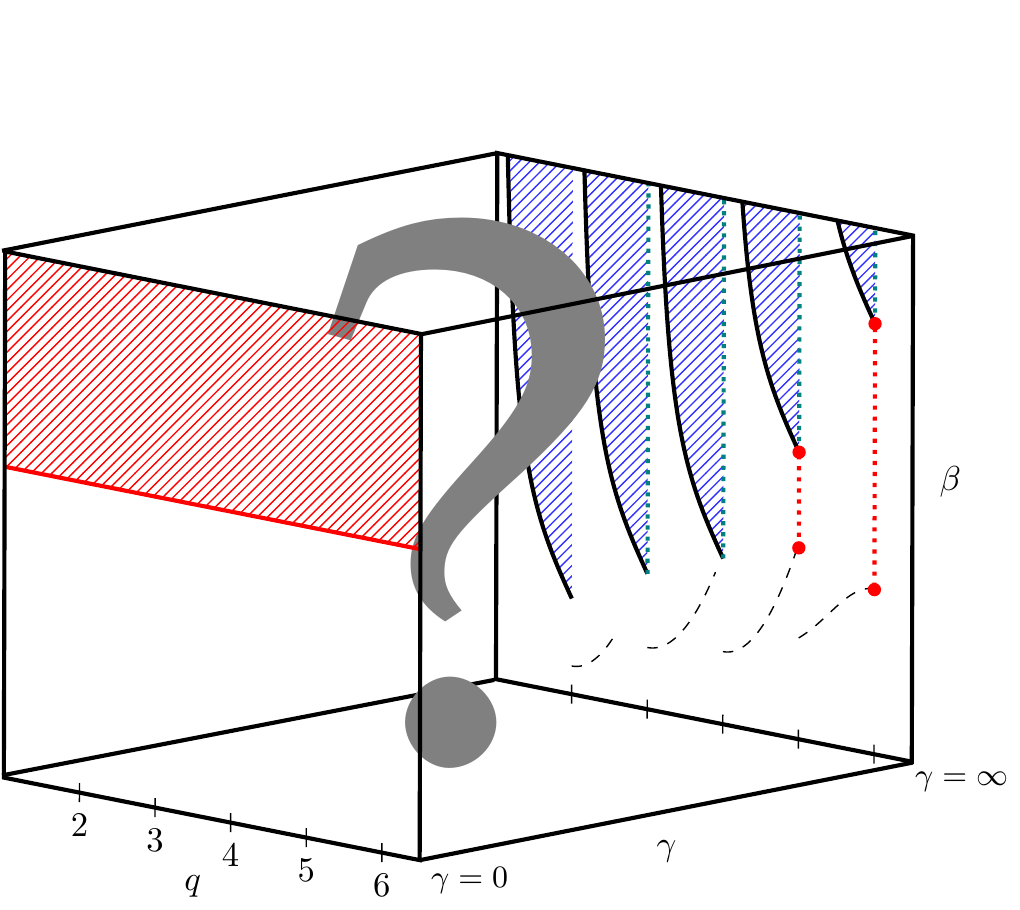}
 \end{figure}

\section{Conclusions}
Tensor Lattice Field Theory is a generic tool to discretize path integral formulations of lattice models. 
For gauge theories, the reformulation can be obtained by a complete integration over the gauge configurations 
and is manifestly gauge-invariant. Microscopic truncations respect symmetries and have interesting critical properties.
Noether theorem can be re-expressed for any kind (continuous, discrete or finite) Abelian symmetry group: for each symmetry, there is a corresponding tensor redundancy. Noise-robust economical implementation of Gauss's law for pure gauge models.
Rydberg atom implementations of scalar QED are expected in the near future. The tensor formulation provides a space-time picture 
that makes clear the building blocks of the 
quantum algorithms and suggest ways to implement them with NISQ devices.

 Acknowledgements: This research was supported in part  by the Dept. of Energy
under Award Numbers DE-SC0010113 and DE-SC0019139. EG was partially supported by the DOE through the Fermilab QuantiSED program in the area of "Intersections of QIS and Theoretical Particle Physics." This manuscript has been authored by Fermi Research Alliance, LLC under Contract No. DE-AC02-07CH11359 with the U.S. Department of Energy, Office of Science, Office of High Energy Physics.



\end{document}